\documentclass[prl]{revtex4}
\usepackage{amssymb}
\begin{document}

\newcommand{\highlite}[1]{\noindent{\bf\underline{#1}}}
\title{\bf Extremal Optimization: Heuristics via
Co-Evolutionary Avalanches}
\author{Stefan Boettcher\\
Dept.~of Physics, Emory University, Atlanta, GA 30322, USA\\
e-mail: stb@physics.emory.edu}
\maketitle
\begin{figure}
\vskip 1.4truein 
\includegraphics{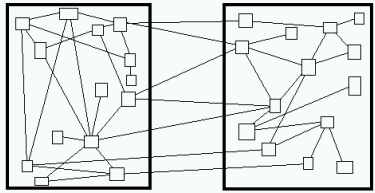}
\caption{Schematic of two integrated circuits with an equal
number of logical gates on each. The graph of connections between
gates (represented by a box at each vertex of the graph) is fixed, but gates can be moved onto either
integrated circuit. Find an equal partition of the gates
such that there is a minimum of connections between the integrated
circuits. How would you verify that a configuration is a global
optimum?}
\label{ic}
\end{figure}

Imagine that you want to design some circuitry for a computer. The
logical function you have to satisfy requires a known network of $n$
interrelated logical gates. Unfortunately, $n$ is too large to put
all the gates on a single integrated circuit, so you have to
partition the gates between separate circuits. Let's assume that
you are forced to place exactly $n/2$ gates each on two integrated
circuits. The connections between gates across the partition are
slow, energy consuming, and heat producing, while the cost
associated with connections inside an integrated circuit are
negligible. So, you have to divide the network of gates such that
the cost function $C$, the number of connections cutting across the
partition, is minimized (see Figure~\ref{ic}). Because a million
computers will be running almost non-stop for 10 years, removing
even one costly connection would be worthwhile.

Fortunately, this (simplified) problem can be mapped onto the
well-known graph bi-partitioning problem.\cite{A+K} In this
problem, the $n$ gates are the vertices of a graph with edges
between two connected gates. Each vertex is a Boolean variable,
with state ``0'' if placed on the left integrated circuit and state
``1'' if placed on the right integrated circuit. Although the graph
of connections is fixed, the vertices can change so that we may
obtain a good partition. Unfortunately, optimizing the equal
partition is NP-hard, that is, the computations needed to find the
global optimum with certainty for even the cleverest algorithm grow
faster than any power of
$n$.\cite{G+J} With even the fastest computers, this
computation would be an unreasonable undertaking for about $n
\gtrsim 100$.

Instead, we can ``search'' the space of all feasible (equal)
partitions $\Omega$. Because the configurations $S\in\Omega$ so far
are unrelated, we need to define a ``neighborhood''
$N(S)\subset\Omega$ for each $S$, a way to proceed from the current
configuration $S$ to some neighboring configuration $S'\in
N(S)$.\cite{Aarts}\cite{Reeves}\cite{Osman} A simple
neighborhood for partitioning is a ``1-exchange:'' for each
$S\in\Omega$, $N(S)$ consists of all $S'\in\Omega$ obtained by
changing a 0-vertex to 1 and a 1-vertex to 0
(to maintain an equal partition). The neighborhood $N$ provides
$\Omega$ with a metric such that the cost function $C(S)$ exhibits
local extrema, like a (high-dimensional) mountain landscape. Then,
moving sequentially ``down-hill'' to better configurations, we
should reach a local minimum very quickly. However, in NP-hard
problems, the number of suboptimal minima of the cost function
grows nearly as fast as the number of configurations, $|\Omega|$,
which here grows like
$|\Omega|={n\choose n/2}\sim e^n$. Thus, in this approach there is no way
to move the system from the current minimum to a better one; it's
like finding the lowest point in a mountainous landscape at night.

In this case, our ``heuristic'' (derived from the Greek word for
``search'') merely produces approximate solutions: local minima of
dubious quality. Can we do better? If we had more time, we could use
an algorithm that makes a small random change of the current
configuration. If the change is big enough, it might take the
system over a ``mountain range'' such that a subsequent descent
would provide a new local minimum. A sequence of these stochastic
updates can only increase our chances to find a better
minimum.\cite{Osman}

A good stochastic search, while inherently slow, succeeds by
controlling the right mixture of descending moves with
``hill-climbing'' perturbations. A particularly elegant stochastic
optimization heuristic is simulated annealing.\cite{Science}
To implement simulated annealing for the graph bi-partitioning
problem, we can adopt the 1-exchange neighborhood by choosing
two vertices at random at each update. If we accept only
1-exchanges that lower the cost, the system converges to a (likely
poor) local minimum and no further improvement is possible for
any pair of vertices that are chosen. In contrast, simulated
annealing allows moves that raise the cost according to the
Metropolis algorithm.\cite{Metro} In each update, a
1-exchange is accepted with probability
$p=\min\{1,e^{-\Delta/T}\}$, where
$\Delta$ is the difference in cost between the new and the old
configuration.

Controlling the ``temperature'' $T$ is crucial for simulated annealing
to succeed: if $T$ is too large, every uphill move is accepted and no
minima are found at all, while for small $T$ only downhill moves are
accepted and the system quickly freezes into a local minimum. Instead,
mimicking the annealing process designed to harden alloys, $T$ is
lowered slowly, allowing simulated annealing to explore the
configuration space landscape widely at high $T$ to reach the largest
``valley.'' In turn, this valley may harbor a correspondingly lower
minimum for the system to freeze into at small $T$. If $T$ is changed
slowly, the Metropolis algorithm ensures detailed
balance and thermodynamic equilibrium (which means that each move
has the same probability as its reverse). 
Because equilibrium systems are well understood, we have an
enormous amount of knowledge to guide simulated annealing.
Theoretically, simulated annealing should always converge to the
global optimum.\cite{G+G} Unfortunately, this convergence
requires vanishingly small decrements of $T$; about as many
updates are needed as in an exhaustive search of
$\Omega$. Short of that, it is a bit of an art to devise a temperature
schedule that balances computational efficiency with the quality of
the minima that are found.\cite{Reeves}

Despite its shortcomings, simulated
annealing is conceptually elegant and often highly
successful for practical problems about which little else is known.
It may even be useful for our graph bi-partitioning
problem.\cite{JohnsonGBP} But could we improve our
procedure by throwing the equilibrium requirement overboard?
Simulated annealing soon collapses for lack of any theoretical
guidance. Although most processes in nature are
out-of-equilibrium, our understanding of these processes
is incomplete. Thus, researchers typically bypass physical
considerations entirely and move to a more abstract conception of
a natural process for inspiration of an optimization procedure.
This is clearly the case for genetic
algorithms.\cite{Holland} The way that a genetic code
evolves and replicates is poorly understood in physical terms. But the basic ingredients by which such a code
selectively optimizes itself is easily transcribed into an
algorithm that operates on gene-like binary encodings of a generic
optimization problem. Yet, without the theoretical guidance that
simulated annealing possesses, the values of the parameters that
control the working of genetic algorithms (mutation rates,
cross-over operators, etc.) are chosen mostly by trial-and-error.
However, with some empirical knowledge, genetic algorithms also
prove to be a powerful optimization procedure with many successful
applications.\cite{Goldberg} 

\medskip \noindent {\bf Emergence and self-organized
criticality}\hfill\break What can we learn by focusing on the
physics of non-equilibrium processes? During the
past decade, some physicists have become interested in systems
exhibiting self-organized criticality, in which complex patterns
emerge without the need to control any
parameters.\cite{Bakbook} For instance, biological
evolution has developed, apparently
by chance, efficient networks in which resources rarely go to
waste.  But species are coupled
in a global comparative process that persistently washes away the
least fit. In this process, unlikely but highly adapted structures
surface inadvertently. Optimal adaptation thus emerges naturally,
without divine intervention, from the dynamics through a selection
{\em against\/} the extremely ``bad.'' In fact, this process
prevents the inflexibility inevitable in a controlled breeding of
the ``good.''

\begin{figure}
\vskip 1.6truein \includegraphics{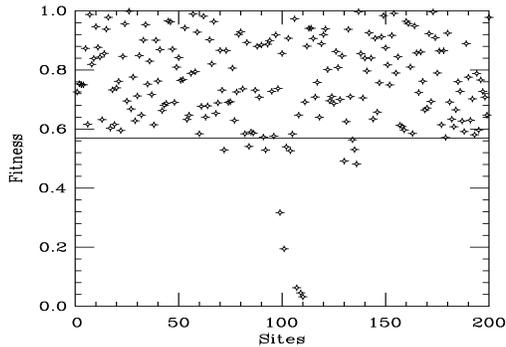}
\caption{Snapshot during the evolution of the Bak-Sneppen model,
showing the fitnesses of a 200-species system. Almost all species
have developed fitnesses above a self-organized threshold
(horizontal line) while a small number of currently active species
have fitnesses below.}
\label{socstate}
\end{figure}

This co-evolutionary process is the basis of the Bak-Sneppen
model, where the high degree of adaptation of most species is
obtained by the elimination of badly adapted ones instead of the
engineering of better ones.\cite{BS} Species in the
Bak-Sneppen model are sites of a lattice, and each is represented
by a value between 0 and 1, indicating its fitness. At each
update, the smallest value (representing the worst adapted
species) is discarded and replaced with a new value drawn randomly
from a flat distribution on
$[0,1]$. Because the change in fitness of one species impacts the
fitness of interrelated species, at each update of the Bak-Sneppen
model, the fitness values on the sites neighboring the smallest
value are replaced with new random numbers as well. After a
certain number of updates, the system organizes itself into a
highly correlated state characteristic of self-organized
criticality.\cite{BTW} In this state, almost all species
have reached a fitness above a certain threshold (see
Figure~\ref{socstate}). But chain reactions, called
``avalanches,'' produce large, non-equilibrium fluctuations in
the configuration of fitness values. The result is that any
possible configuration is accessible.

\medskip \noindent {\bf Extremal optimization}\hfill\break
The extremal dynamics of the Bak-Sneppen model can be converted
into an optimization algorithm that we call extremal
optimization.\cite{GECCO} Attractive
features of the model include the following:

\begin{itemize}
\item It is straightforward to relate the sum of all
fitnesses to the cost function of the system.
\item In the self-organized critical state to which the system
inevitably evolves, almost all species have a much better than
random fitness (see Figure~\ref{socstate}).
\item Most species preserve a good fitness for long times unless
they are connected to poorly adapted species, providing the system
with a long memory.\cite{BoPa1}
\item The system retains a potential for large, hill-climbing
fluctuations at any stage.
\item Aside from picking the worst species for an update, the
model accomplishes these features without a single control
parameter.
\end{itemize}

To be precise, we define $S=(x_1,\ldots,x_n)\in\Omega$ to be a
configuration of the $n$ variables $x_i$ in an optimization
problem. For instance, in the graph bi-partitioning problem the
variables $x_i$ are the vertices, which can take on the values 0 or
1; a configuration $S$ is one possible arrangement of $n/2$
0's and $n/2$ 1's. The cost function $C(S)$ simply counts the
number of bad edges that connect a 0 with a 1 in $S$. Finally, we
define a neighborhood $N(S)$ that maps $S\to S'\in
N(S)\subset\Omega$ to facilitate a local search, like the
1-exchange for the graph bi-partitioning problem.

Extremal
optimization performs a search through sequential changes on a
single configuration $S\in\Omega$. The cost $C(S)$ is assumed to
consist of the individual cost contributions $\lambda_i$ for each
variable
$x_i$, which correspond to the fitness values in the
Bak-Sneppen model above. Typically, the fitness $\lambda_i$ of
variable $x_i$ depends on its state in relation to other variables
to which $x_i$ is connected. Ideally, it is possible to write
the cost function as 
\begin{eqnarray}
C(S)=\sum_{i=1}^n \lambda_i.
\label{costeq}
\end{eqnarray}
For example, in the graph bi-partitioning problem
Eq.~(\ref{costeq}) is satisfied, if we attribute to each vertex
$x_i$ a local cost $\lambda_i=b_i/2$, where
$b_i$ is the number of bad edges of $x_i$, whose cost is equally shared
with the vertices on the other end of those edges. 

For minimization problems, extremal optimization proceeds as
follows: 
\begin{center}
\begin{enumerate}
\item Initialize a configuration $S$ at will and set $S_{\rm
best}=S$.
\item For the current configuration $S$,
\label{EOupdate}
\begin{enumerate}
\item evaluate $\lambda_i$ for each variable $x_i$;
\label{evaluate}
\item find $j$ with $\lambda_j\geq\lambda_i$ for all $i$, that is,
$x_j$ has the worst fitness;
\label{sort}
\item choose a random $S'\in N(S)$ such that $x_j$ must change;
\label{neighbor}
\item if $C(S')<C(S_{\rm best})$, store $S_{\rm best}=S'$;
\item accept $S\leftarrow S'$ {\em always, independent\/} of
$\Delta=C(S')-C(S)$.
\label{alwaysmove}
\end{enumerate}
\item Repeat at step~(\ref{EOupdate}) as long as desired.
\item Return $S_{\rm best}$ and $C(S_{\rm best})$.\\
\end{enumerate}
\end{center}

\begin{figure}
\vskip 1.5truein \includegraphics{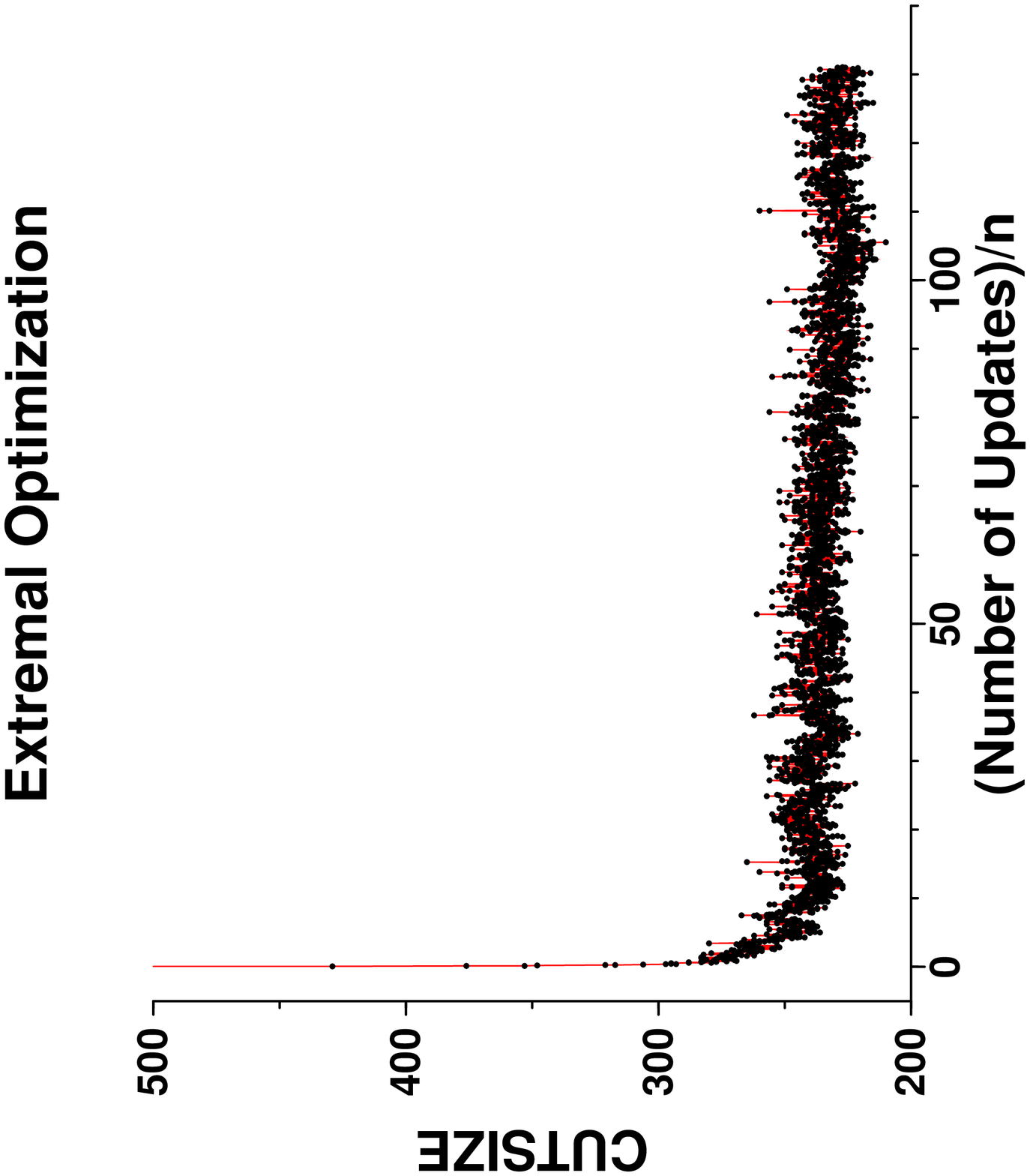} \includegraphics{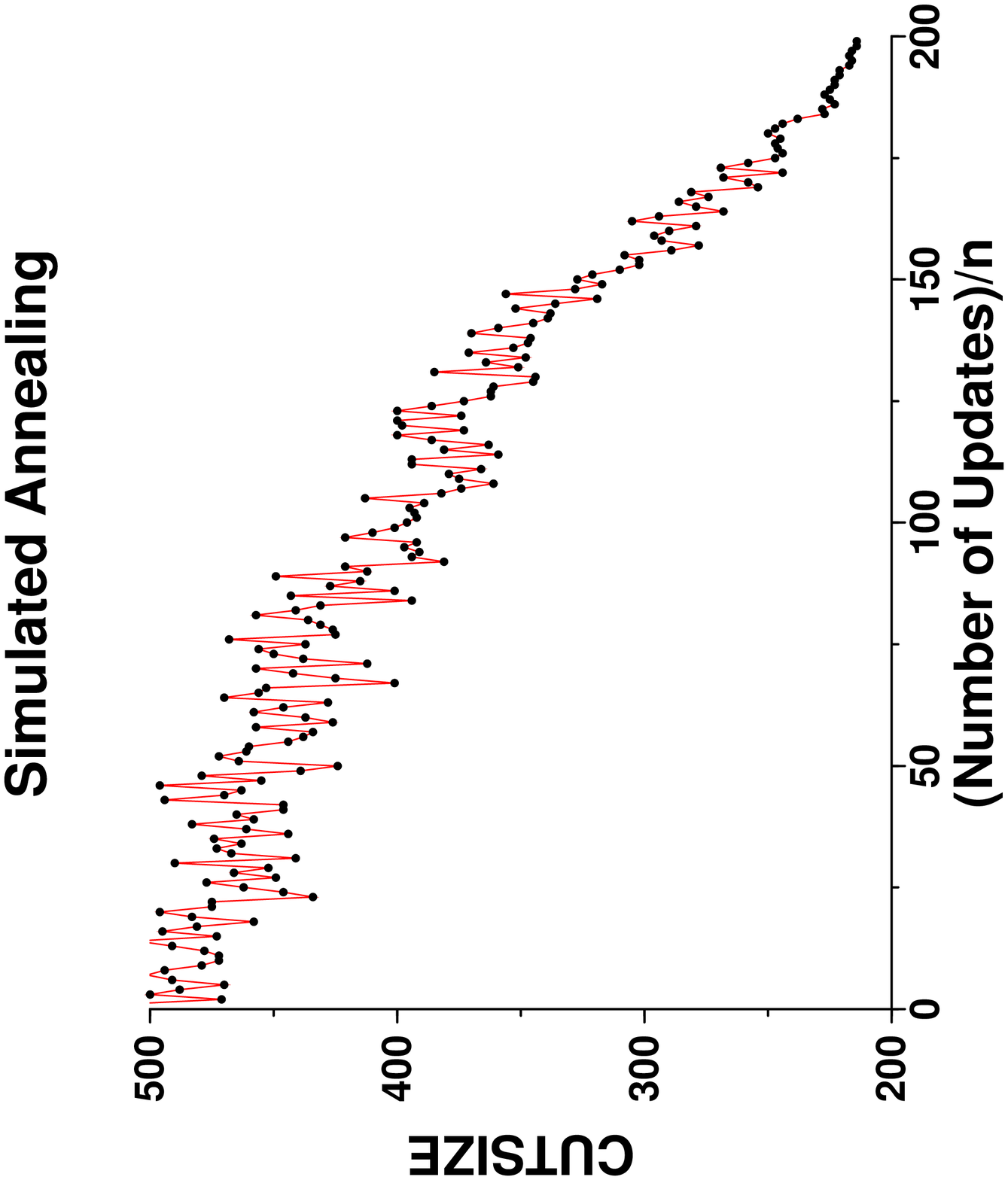}
\caption{Evolution of the cost function $C(S)$ during a typical run of (a)
extremal
optimization and (b) simulated
annealing for the $n=500$ random graph $G_{500}$ introduced in
Ref.~\protect\cite{JohnsonGBP}. The lowest cost ever found for
$G_{500}$ is 206 (see Figure~\protect\ref{fig9}). In contrast to
simulated
annealing, which has large fluctuations in early stages of the run and
then converges much later, extremal optimization quickly
approaches a stage where broadly distributed fluctuations allow it
to scale barriers and probe many local minima.}
\label{runtime}
\end{figure}

A typical run of this implementation of extremal optimization for
the graph bi-partitioning problem on an $n=500$ random graph is
shown in Figure~\ref{runtime}a.

The most apparent distinction between extremal optimization and other
methods is the need to define local cost contributions for each
variable, instead of merely a global cost. Extremal optimization's
capability appears to derive from its ability to access this local
information directly. Extremal optimization's ranking of fitnesses
required for step~(\ref{sort}) 
above superficially appears like the rankings of possible moves in some
versions of simulated annealing and other
heuristics.\cite{Reeves} There, moves are evaluated by their
{\em anticipated outcome}, while extremal optimization's fitnesses
reflect the {\em current} configuration $S$ without biasing the
outcome. As a comparison of Figures~\ref{runtime}a and 3b
demonstrates, a sequence of these moves allows for much larger
than equilibrium fluctuations in $C(S)$.

Although similarly motivated, genetic
algorithms\cite{Holland}\cite{Goldberg} and extremal
optimization algorithms have hardly anything in common. Genetic
algorithms mimic evolution on the level of genes, and keep track of
entire gene pools of configurations from which to select and breed an
improved generation of solutions. In comparison, extremal
optimization, based on evolutionary competition at the
phenomenological level of species, operates only on a single
configuration, with improvements achieved merely by the elimination
of bad fitness values. Extremal optimization and simulated
annealing perform a local search but in genetic algorithms
cross-over operators perform global exchanges.

\begin{figure}
\vspace{1.35in} \includegraphics{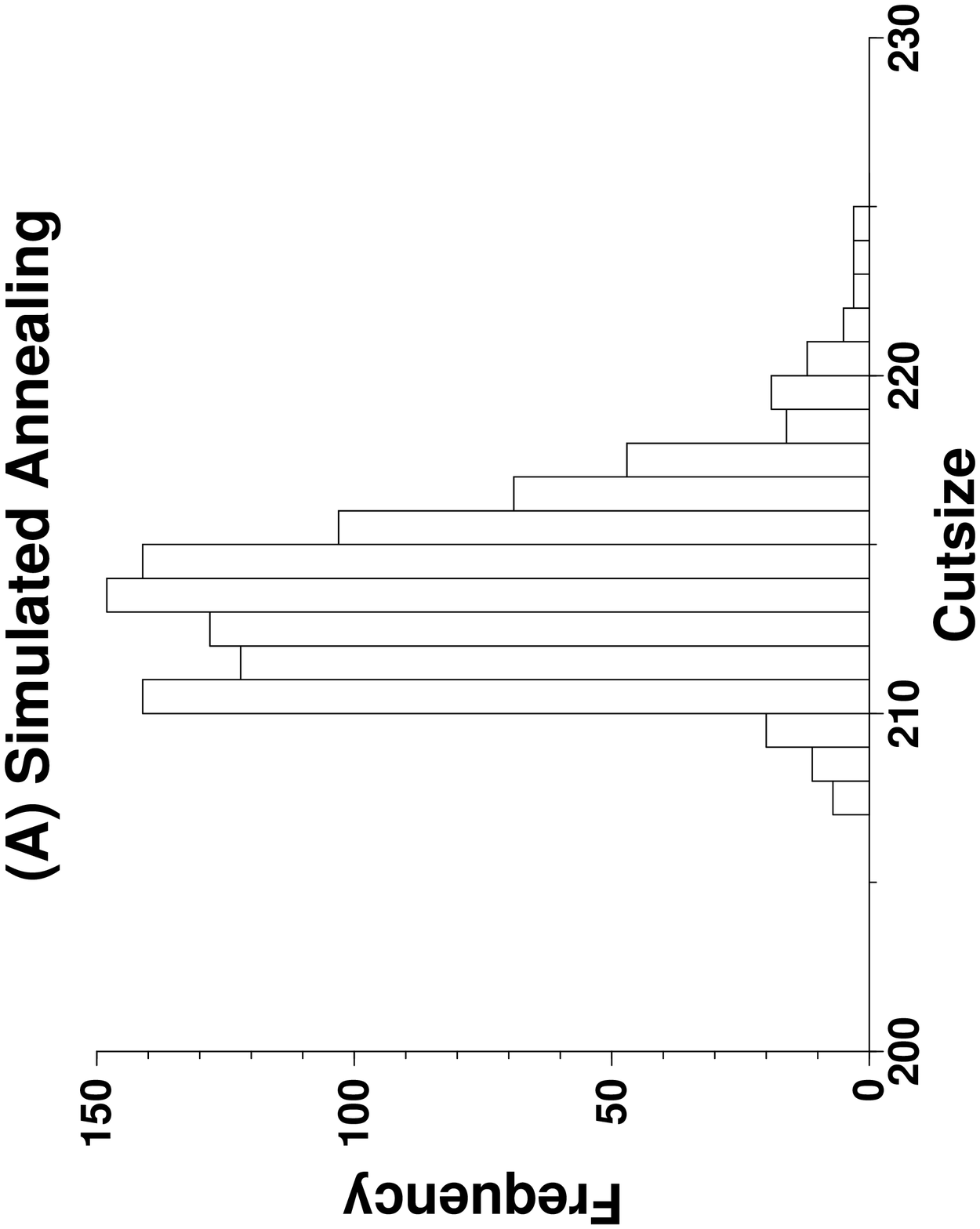} \includegraphics{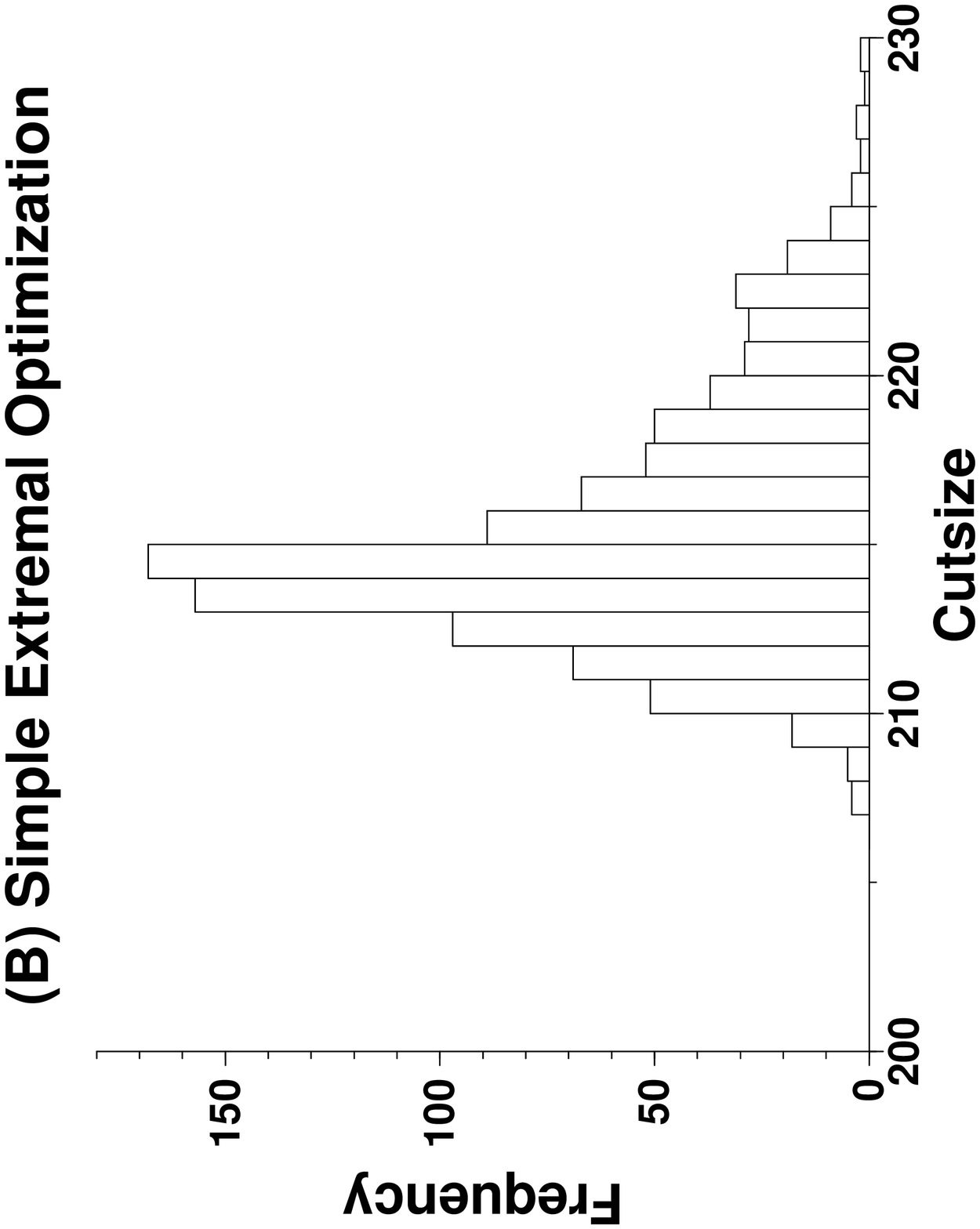} \includegraphics{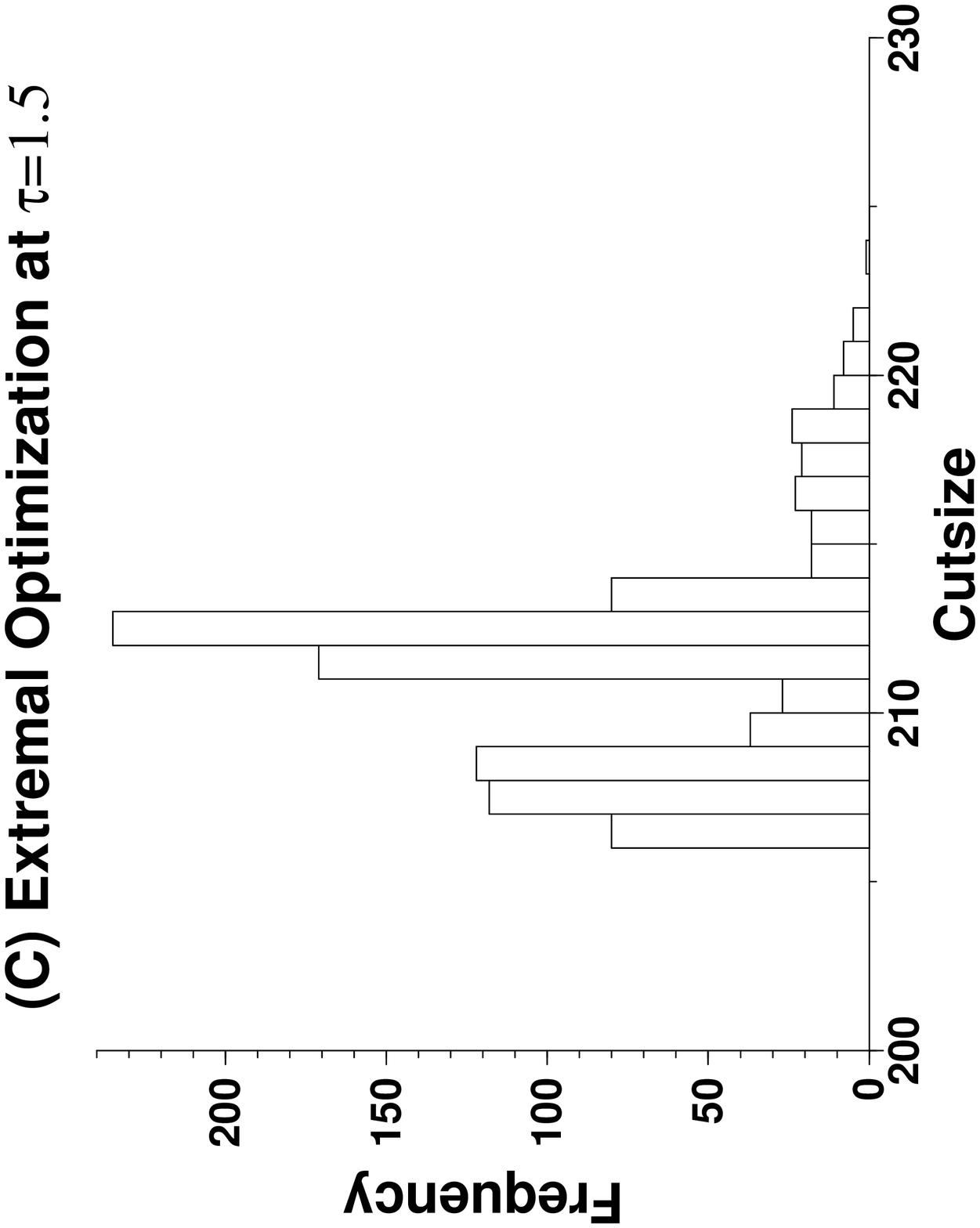}
\caption{Comparison of 1000-run trials using various optimization
methods on $G_{500}$.\protect\cite{JohnsonGBP} The
histograms give the frequency with which a particular cost has
been obtained during the trial runs for (a) simulated
annealing, (b) basic extremal
optimization, and (c) for $\tau$-extremal
optimization with
$\tau=1.5$. The best cost ever found for this graph is 206. This
result appeared only once over the 1000 simulated
annealing runs, but 80 times for $\tau$-extremal optimization.}
\label{fig9} 
\end{figure}

\medskip \noindent {\bf Simple extremal optimization application to 
graph partitioning}\hfill\break
Following the example of
Ref.~\cite{JohnsonGBP} (see Figure~9), we tested early
implementations of extremal optimization\cite{BoPe1} on their
$n=500$ random graph $G_{500}$. This version is based on a 1-exchange
between the worst vertex [step~(\ref{sort})] and one randomly chosen
vertex of the opposite state [step~(\ref{neighbor})]. In a 1000-run
sample from different random initial conditions, we determined the
frequency of solutions (see Figure~\ref{fig9}). For comparison, we
have also implemented the simulated annealing algorithm as given in
Ref.~\cite{JohnsonGBP} on the same data structure used by our extremal
optimization program. We have used run times for extremal optimization
that are about three times longer than the time it took for simulated
annealing to freeze, because extremal optimization still yielded
significant gains. We checked that neither the best-of-three runs of
simulated annealing, or a three times longer temperature schedule,
improved the simulated annealing results significantly. Although the
basic, parameter-free version of extremal optimization considered so
far is already competitive, the best results are obtained by
$\tau$-extremal optimization, which we now discuss.

\medskip \noindent {\bf The $\tau$-extremal optimization 
implementation}\hfill\break The $\tau$-extremal optimization 
implementation is a modification of extremal optimization
that improves results and avoids dead ends that occur in some
implementations at the expense of introducing a single
parameter.\cite{BoPe1} In general, the
implementation of $\tau$-extremal optimization proceeds as
follows. Rank all the variables $x_i$ according to their fitness
$\lambda_i$, that is, find a permutation $\Pi$ of the labels $i$
such that 
\begin{eqnarray}
\lambda_{\Pi(1)}\geq\lambda_{\Pi(2)}\geq\ldots\geq\lambda_{\Pi(n)}.
\end{eqnarray}
The worst variable $x_j$ [see step~(\ref{sort})] is of rank 1, $j=\Pi(1)$, and
the best variable is of rank $n$. Consider a probability distribution
over the {\em ranks\/} $k$,
\begin{eqnarray}
P_k\propto k^{-\tau},\qquad 1\leq k\leq n,
\label{taueq}
\end{eqnarray}
for a fixed value of the parameter $\tau$. At each update, for each
independent variable $x$ to be moved, select distinct ranks
$k_1,k_2,\ldots$ according to $P_k$. Then, execute
step~(\ref{neighbor}) such that all $x_{i_1},x_{i_2},\ldots$ with
$i_1=\Pi(k_1),~i_2=\Pi(k_2),\ldots$ change. For example, in the
bi-partitioning problem, we choose {\em both\/} variables in the
1-exchange according to $P_k$, instead of the worst and a random
one. Although the worst variable of rank $i=1$ will be chosen most
often, sometimes (much) higher ranks will be updated instead. In fact,
the choice of a power-law distribution for $P_k$ (instead of, say,
an exponential distribution with a cut-off scale excluding high
ranks) ensures that no rank gets excluded from further evolution
while maintaining a bias against variables with bad fitness.

Clearly, for $\tau=0$, $\tau$-extremal optimization is exactly a
random walk through $\Omega$. Conversely, for $\tau\to\infty$, the
process approaches a deterministic local search, only swapping the
lowest-ranked variables, and is bound to reach a dead end. 
Indeed, tests of both $\tau=0$ and $\tau=\infty$ yield terrible
results. In the graph bi-partitioning problem, we obtained our best
solutions for $\tau$ in the range 1.4--1.6. Clearly, we have ``tuned''
away from the philosophy of the Bak-Sneppen model by inserting a
single parameter for the sake of better results. To be successful,
{\em every} heuristic has to allow for some adjustments to a
particular problem.

\begin{table}[b!]
\caption{ Best costs (and allowed run time (in seconds)) for
a testbed of large graphs. The value of $n$ is given for
each graph. Genetic algorithms results are the best
reported\protect\cite{MF1} (using a 300\,MHz CPU).
The $\tau$-extremal optimization results
are from our runs (200\,MHz). Comparison data for three of the
large graphs are due to results from heuristics in
Ref.~\protect\cite{HL} (50\,MHz). METIS is a partitioning
program based on hierarchical reduction instead of local
search,\protect\cite{METIS} and yields extremely
fast, deterministic results (200\,MHz). }
\begin{tabular}{ll|rr|rr|rr|rr}
\hline Large Graph& & GA& & $\tau$-EO&&{\protect\cite{HL}}& & METIS&
\\ \hline {\em Hammond\/}& ($n=4720$) & 90 & (1s) & 90 & (42s) & 97 &
(8s) & 92 & (0s) \\  {\em Barth5\/} & ($n=15606$) & 139 & (44s) & 139
& (64s) & 146 & (28s) & 151 & (0.5s)\\  {\em Brack2\/} & ($n=62632$) &
731 & (255s) & 731 & (12s) & ---& & 758 & (4s) \\  {\em Ocean\/} &
($n=143437$) & 464 & (1200s) & 464 & (200s) & 499 & (38s) & 478 &
(6s)\\ \hline
\end{tabular}
\label{tab1}
\end{table}

Table~\ref{tab1} summarizes our $\tau$-extremal optimization results
on some well-studied instances of graphs with large $n$, using
$\tau=1.4$ and best-of-10 runs. We obtained initial configurations
from a simple clustering algorithm.\cite{BoPe1} Then, we
choose the number of updates $t$ to obtain reliable results that do
not change much with $t$. The choice of $t$ varied with the
particularities of each graph, from $t=2n$ to $t=200n$, and the
reported run times are of course influenced by the value of $t$. It is
worth noting, though, that extremal optimization's average performance
has been varied. For instance, half of the runs on the graph known
as Brack2 returned costs near 731, but the other half returned
costs above 2000. This may be a product of an unusual structure in
this particular graph. A systematic study of
extremal optimization in comparison with simulated annealing by
averaging over many graphs of increasing size is given in
Ref.~\cite{EOperc}.

\medskip \noindent {\bf Other extremal
optimization implementations}\hfill\break To demonstrate the
generality of extremal optimization, we are currently studying its
implementation for other NP-hard optimization problems such as graph
coloring ($K$-COL), satisfiability ($K$-SAT), and spin glass
Hamiltonians. In $K$-COL, given $K$ different colors to label the
vertices of a graph, we need to find a coloring that minimizes the
number of edges connecting vertices of identical
color.\cite{JohnsonCOL} A definition of fitness is as
obvious as it was for the graph bi-partitioning problem: for each
vertex $x_i$ simply count the number $b_i$ of equally colored vertices
connected to it; setting $\lambda_i=b_i/2$ again satisfies
Eq.~(\ref{costeq}). The lack of a global constraint as for the
graph bi-partitioning problem allows us to define a neighborhood
by changing the state of only one, the worst, variable.
However, this definition results in a deterministic search that
quickly reaches a dead end. But a
$\tau$-extremal optimization implementation picking a single
variable with
$\tau\approx2.7$ seems to work best for the graphs we have studied.

\begin{figure}
\vskip 1.5truein \includegraphics{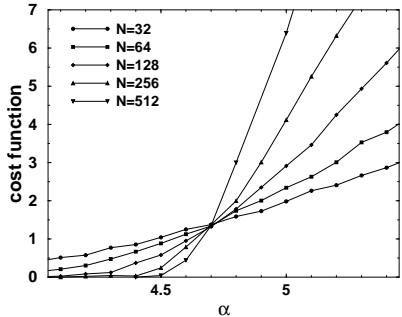}
\caption{Plot of the average cost as a function of the average vertex degree
$\alpha$ for 3-COL of random graphs. We have generated $2,300$, $500$, and
$280$ instances for $n=32$, 64, and 128, respectively, at each
value of
$c$. The prediction for the critical point (where the cost functions 
intersect) of
$\alpha_{\rm crit}(3)\approx4.73$ is indicated by a vertical line.}
\label{transition}
\end{figure}

With this algorithm we have studied the ``phase transition'' of
$K$-COL,\cite{BIP} ``where the really hard instances
are.''\cite{Cheeseman} This transition arises as a function of
the average vertex degree $\alpha$ for certain types of graphs. (The
``degree'' of a vertex is the number of edges emanating from it and
may vary between vertices of a graph.) If $\alpha$ is small, almost
all vertices have fewer than $K$ neighbors, coloring becomes trivial
and the optimal solution has zero cost for almost all graph
instances. But around a critical value $\alpha_{\rm crit}(K)$, the
cost becomes positive, with an ever sharper transition for
$n\to\infty$. If we average the best solutions that extremal
optimization finds over many instances of random
graphs,\cite{Bollobas} we can show, for example, that
$\alpha_{\rm crit}(3)\approx4.73$ (see Figure~\ref{transition}).
The relationship between phase transitions, which occur in many
NP-hard problems, and computational complexity is evolving into
one of the hot topics in computer science.\cite{Cheeseman}
\cite{AI}\cite{Monasson}\cite{BIP}\cite{Machta} 
In this interesting regime extremal optimization's large fluctuations
appear to have the edge over simulated annealing's equilibrium
requirements.\cite{EOperc}

Because extremal optimization is fairly new, there are many questions
yet to be answered. It will not be difficult for the interested reader
to think of research projects that, for example, compare extremal
optimization to other methods. Clearly, like any other optimization
method, extremal optimization will not be competitive for some
problems (unfortunately, the traveling salesman problem seems to be
one example.\cite{BoPe1}) But it can't hurt to have more
alternative methods to choose from for tackling hard optimization
problems.

\medskip \noindent {\bf Acknowledgements}\hfill\break
I would like to thank Allon Percus with whom I developed extremal 
optimization, and the Research Committee at Emory University for
support.

\medskip \noindent {\bf Suggestions for
Further Study}
\begin{enumerate}
\item A simple model of a glass consists of a
$d$-dimensional hypercubic lattice with a spin variable
$\sigma_i\in [-1,1]$ placed on each site $i$, $1\leq i\leq
n=L^d$.\cite{MPV} Every spin is connected to each of its
nearest neighbors $j$ via a fixed bond variable $J_{i,j}$ drawn randomly from
$[-1,1]$. Spins may be coupled to an arbitrary external field
$h_i$. The cost function to be minimized is the Hamiltonian
\begin{eqnarray}
C(S)=H(\sigma_1,\ldots,\sigma_n)=-{1\over2}\sum_i\sum_jJ_{i,j}
\sigma_i\sigma_j-\sum_i\sigma_i h_i.
\label{spineq}
\end{eqnarray}
Arranging the spins into an optimal, lowest energy configuration is
hard due to frustration.\cite{MPV} In fact, for $d>2$ the
problem is NP-hard.\cite{Barahona}
\begin{enumerate}
\item Find a definition of $\lambda_i$ for each spin variable such
that Eq.~(\ref{costeq}) is satisfied.
\item Define a simple neighborhood $N$ for this problem using only single
spin flips. In this case, the basic extremal optimization implementation is
equivalent to $\tau$-extremal optimization for a certain value of $\tau$.
(Why is this not the case for the basic partitioning implemetation?) Do you
think that the basic implementation would be very successful?
\item Implement an algorithm for this spin glass in $d=3$ for $h_i=0$, using the hash table from the
Problem~2. A single run should have at least $t>n$
updates (why?); to be save set $t=100n$. Find a good value
for
$\tau$. Does it depend on $n$ or $t$? For comparison, the cost
function has been found for genetic algorithms to scale as
$C(S_{\rm best})\sim -1.786n$ for
$n=3^3\ldots14^3$.\cite{Pal}
\end{enumerate}

\medskip
\item The $\tau$-extremal
optimization algorithm described in the text requires a perfect
ordering of the $\lambda_i$ ($1\leq i\leq n$), which would produce a
factor of $n\ln n$ for the computational time. In practice, it is
sufficient to order the $\lambda_i$ somewhat. (Most likely many
$\lambda_i$ will be degenerate.)
\begin{enumerate}
\item Devise a $\tau$-extremal
optimization implementation in which the $\lambda_i$ are sorted on a
binary tree only (time factor $\ln n$), where the $\lambda_i$ are
picked such that Eq.~(\ref{taueq}) is roughly approximated? (For our
humble attempt, see Ref.~\cite{BoPe1}.)
\item For some problems even the use of a hash
table\cite{K+P} with a constant time factor may be useful.
For the spin glass problem in Problem~1, exploit the degeneracies between individual fitnesses to give a sorting algorithm
using a hash table that leaves the $\lambda_i$ perfectly ordered.
\end{enumerate}
\end{enumerate}

\end{document}